\documentclass[aps,nofootinbib,twocolumn,showpacs,superscriptaddress,prc]{revtex4}
\date{\today}
\usepackage{graphicx}
\usepackage{color}
\usepackage{amsmath}
\usepackage{amsfonts}
\usepackage{amssymb}

\begin{document}

\title{Energy expressions for $n=3$ and 4 systems in a single-$j$ shell}

\author{Chong Qi}
\thanks{Email: chongq@kth.se}
\affiliation{KTH (Royal Institute of Technology), Alba Nova University Center,
SE-10691 Stockholm, Sweden}

\begin{abstract}
For systems with three and four fermions within a single-$j$ shell, analytical expressions for the state energies
are presented from a decomposition of the angular momentum. In some important cases the expressions acquire a very simple form. The expression may help us in understanding the structure of isomeric states. The decomposition also makes it possible to construct the algebraic condition for conservation of seniority.
\end{abstract}


\maketitle
The developments in experimental techniques have made it possible to study the structure of $N\approx Z$ nuclei far from the stability line.
For $N\approx Z$ nuclei just above $^{40}$Ca and immediately below $^{100}$Sn, the structure of most low-lying states is dominated by the coupling of  few valence particles (or holes) moving in a single-$j$ shell-model orbit.
In a single-$j$ shell, a state with angular momentum $I$ may be constructed for which the
expectation value of a two-body interaction can be written as,
\begin{equation}
E_I=C_J^IV_J,
\end{equation}
where $V_J=\langle j^2;J|\hat{V}|j^2;J\rangle$ are two-body matrix elements and $C^I_J$ are corresponding expansion coefficients (The constant contribution from the single particle energy has been neglected for simplicity).  If isopsin symmetry is conserved in the two-body interaction $\hat{V}$, then one has a total number of $2j+1$ matrix elements being coupled to spin values $J=0$ to $2j$~\cite{Talmi93,Talmi03}.  The expansion coefficient $C^I_J$ gives the number of pairs with angular momentum $J$~\cite{Zam05}.
Some trivial examples are the
two-nucleon system, for which $E_J=V_J$, and the case of a closed single-$j$ shell, for which $E_J=\sum_{J}(2J+1)V_J$~\cite{Talmi93,Talmi03}.

The total number of pairs with all spins $J$ is given by~\cite{Zam05,Moya03},
\begin{equation}
\sum_{J}C^I_J=n(n-1)/2,
\end{equation}
and
\begin{equation}
\sum_{J,{\rm odd}}C^I_J=\frac{1}{2}\left[\frac{n}{2}\left(\frac{n}{2}+1\right)-T(T+1)\right],
\end{equation}
where $n$ is the total number of nucleons and $T$ is the total isospin quantum number of the system.

In this work systems with $n=3$ and 4 nucleons are analyzed and expressions for the number of pairs with any total angular momentum and isospin are derived in terms of angular momentum coupling coefficients. Simplified relations are also derived when possible. For earlier works based on coefficients of fractional parentage calculations, see Refs.~\cite{Zam05,Moya03,Zam07}. In Ref.~\cite{Zam05}, the number of $J=0$ pairs in the $0^+$ states of the $n=4$ nucleus $^{44}$Ti is obtained. We will start from the so-called multistep shell model~\cite{Liotta81}. It employs correlated two-particle bases as building blocks and was first proposed as a truncation of the exact shell-model approach~\cite{Liotta81} (The generalization of this model to many-body systems is given in Refs.~\cite{Liotta82,Liotta82b}). When restricted to a single-$j$ shell, the correlated basis with spin $J$ is closely related to the corresponding matrix element $V_J$, which makes the multistep shell model a very convenient tool to study the energy expressions of nuclear systems. 


For a $n=4$ system with two protons and two neutrons, the system can be decomposed into proton and neutron blocks. The wave function of a given state with total angular momentum $I$ can
be written as~\cite{Ryd90},
\begin{equation}\label{4pn}
|\Psi_I\rangle=\sum_{J_{p}, J_n} X_I(J_pJ_n) | j_{\pi}^2(J_p)j_{\nu}^2(J_n);I \rangle,
\end{equation}
where $X_I(J_pJ_n)$ is the amplitude of the four-body wave function and $J_p$ and $J_n$ are even numbers denoting the angular momenta of the proton and neutron pairs, respectively.

The four nucleons can couple to spin $I=0$ to $2(2j-1)$ and isospin $T=0$, 1 and $2$. The $T=2$ state is the double analog of the system of four identical nucleons. 

The dimension of a given state $I$ will be denoted by $D_I(j)$, while for the number of states 
with a given isospin $T$  the notation $D_I^T(j)$ will be used. Recently, there have been efforts 
to obtain algebraic formulas for the state dimension in a single-$j$ shell~\cite{Zhao05,Zam05a,Talmi}. 
For $I=0$ and $n=4$ these quantities are relatively easy to evaluate. The dimension equals to the number of $T=1$ states in the two-body system, i.e., $D_0(j)=(2j+1)/2$. 
For spin zero states with $T=2$ one has $D_0^2(j)=[(2j+3)/6]=[(j+2)/3]$ ($[n]$ denotes the largest integer not exceeding $n$)~\cite{Gino93,Zam05b}. For the $T=0$ states it is $D^0_0(j)=D_0(j)-D_0^2(j)=(2j+1)/2-[(2j+3)/6]$.
That is, for $j=3/2$ to $7/2$ one gets $D_0^0(j)=(2j-1)/2$ and $D_0^2(j)=1$. For $j=9/2$ to $13/2$, 
we have $D_0^0(j)=(2j-3)/2$ and $D_0^2(j)=2$.
For $I=0$, there is no $T=1$ state.

In the bases of Eq.~(\ref{4pn}), the matrix elements of the Hamiltonian can be expressed as linear combinations of  two-body interactions $V_J$ as,
\begin{eqnarray}\label{4h}
\nonumber \langle  j_{\pi}^2(J_p)j_{\nu}^2(J_n);I |\hat{V}| j_{\pi}^2(J_p')j_{\nu}^2(J_n');I \rangle\\
= (V_{J_p}+V_{J_n})\delta_{J_pJ_p'}\delta_{J_nJ_n'}
+\sum_JM^I_J(J_pJ_n;J_p'J_n')V_J,
\end{eqnarray}
where the spin $J$ can take both even and odd values ($J=0$ to $2j$).
The matrix $M$ is given as~\cite{Ryd90},
\begin{eqnarray}
\nonumber M^I_J(J_pJ_n;J_p'J_n')=\sum_{\lambda}4\hat{J}_p\hat{J}_n\hat{J}_p'\hat{J}_n'\hat{J}^2\hat{\lambda}^2\left\{
\begin{array}{ccc}
 J_p & J_n & I \\
 \lambda & j & j \\
\end{array}\right\}\\
\times\left\{\begin{array}{ccc}
 J_p' & J_n' & I \\
 \lambda & j & j \\
\end{array}
\right\}\left\{\begin{array}{ccc}
 j & j & J \\
 \lambda & j & J_n \\
\end{array}
\right\}\left\{\begin{array}{ccc}
 j & j & J \\
 \lambda & j & J_n' \\
\end{array}
\right\},
\end{eqnarray}
where $\hat{J}=\sqrt{2J+1}$ and $\lambda$ and $j$ are half integers. Both the Hamiltonian matrix and $M$ are symmetric since we have $M^I_J(J_pJ_n;J_p'J_n')=M^I_J(J_p'J_n';J_pJ_n)$.

Direct diagonalization of the Hamiltonian matrix (Eq.~(\ref{4h})) would give eigen energies as nonlinear combinations of two-body interactions $V_J$. But still the eigen energy can be re-expressed in the linear expansion form of Eq.~(1).
Generally, the expansion coefficients for a state with total angular momentum $I$ of the $n=4$ system can be written as,
\begin{eqnarray}\label{num4}
\nonumber C^I_J=|X_I(J_pJ_n)|^2(\delta_{J_pJ}+\delta_{J_nJ})\\
+\sum_{J_pJ_n;J_p'J_n'}X_I(J_pJ_n)M^I_J(J_pJ_n;J_p'J_n')X_I(J_p'J_n'),
\end{eqnarray}
where the first and second terms in the right-hand side give contributions from identical nucleon pairs and proton-neutron pairs, respectively. 
The wave function amplitudes $X$ as well as the coefficients $C$ can be calculated numerically by diagonalizing the Hamiltonian matrix.
The thus calculated  coefficients are usually irrational numbers and depend on the interactions employed~\cite{Zam05}.

\begin{figure}\centering
\includegraphics[width=8.5cm]{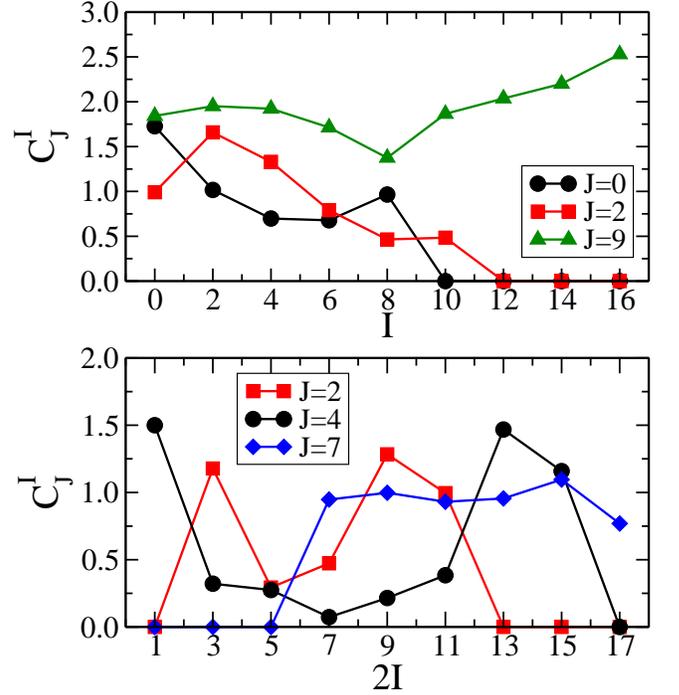}
\caption{(Color online)
Upper: expansion coefficients $C^I_J$ as a function of spin $I$ for yrast states of $^{96}$Cd in the $0g_{9/2}$ shell; Lower: same as the upper panel but for those of $^{43}$Sc (and $^{43}$Ti) in the $0f_{7/2}$ shell.\label{fig1}}
\end{figure}

With the expression given above, calculations have been done to explore the energy expressions in several single-$j$ shells. As examples,  in the upper panel of Fig.~\ref{fig1} coefficients $C^I_J$ are plotted as a function of spin $I$ for yrast states of $^{96}$Cd. Calculations are done in the $0g_{9/2}$ shell with two-body interactions extracted from Ref.~\cite{js}.

Eq.~(\ref{num4}) is valid for all spins and isospins. But in some cases, the expression can be greatly simplified based on the symmetric properties of the $6j$ symbols. 
It is easily seen that for $I=0$ we have ${\rm rank}(M)=1$ and,
\begin{equation}
M^0_J(J_pJ_n;J_p'J_n')=\alpha_J({J_p})\alpha_J({J_p'})\delta_{J_pJ_n}\delta_{J_p'J_n'},
\end{equation}
where
\begin{eqnarray}
\nonumber \alpha_J(J_p)&=&2\hat{J}_p^2\hat{J}\hat{j}\left\{
\begin{array}{ccc}
 J_p & J_p & 0 \\
j & j & j \\
\end{array}\right\}
\left\{\begin{array}{ccc}
 j & j & J \\
j & j & J_p \\
\end{array}
\right\}\\
&=&-2\hat{J}_p\hat{J}
\left\{\begin{array}{ccc}
 j & j & J \\
j & j & J_p \\
\end{array}
\right\}.
\end{eqnarray}
For the number of $J=0$ pairs in system with any angular momentum $I$ one gets,
\begin{eqnarray}
\nonumber M^I_0(J_pJ_n;J_p'J_n')&=&4\hat{J}_p\hat{J}_n\hat{J}_p'\hat{J}_n'/\hat{j}^2\left\{
\begin{array}{ccc}
 J_p & J_n & I \\
j & j & j \\
\end{array}\right\}\\
&&\times\left\{\begin{array}{ccc}
 J_p' & J_n' & I \\
 j & j & j \\
\end{array}
\right\}.
\end{eqnarray}
For $J=0$ and $I=0$, the expression can be even simpler. It is $\alpha_0(J_p)=2\hat{J_p}/\hat{j}^2$ and
\begin{equation}
C^0_0=2|X_0(00)|^2+\frac{4}{\hat{j}^4}\sum_{J_pJ_p'}\hat{J}_p\hat{J}_p'X_0(J_pJ_p)X_0(J_p'J_p').
\end{equation}

Another interesting case is the state with the maximum spin of $I=4j-2$ and dimension $D=1$. For this unique state we have $J_p=J_n=2j-1$ and $|X_I(J_pJ_n)|=1$.  Inserting these into Eq.~(\ref{num4}), it can be easily seen that we have $C^I_J=0$ and 3 for the number of pairs with $J<2j-2$ and $J=2j-1$, respectively. For that of $J=2j-2$ we have,
\begin{eqnarray}
\nonumber C^{I=4j-2}_{J=2j-2}&=&12(4j-1)^2(4j-3)(2j-1)\\
\nonumber &&\times\left|\left\{\begin{array}{ccc}
 2j-1 & 2j-1 & 4j-2 \\
3j-2 & j & j \\
\end{array}
\right\}\right.\\
&& \times\left.\left\{\begin{array}{ccc}
j & j & 2j-2 \\
3j-2 & j & 2j-1 \\
\end{array}
\right\}\right|^2.
\end{eqnarray}
The number of spin $2j$ pairs is determined as $C^{I=4j-2}_{J=2j}=3-C^{I=4j-2}_{J=2j-2}$.
These numbers of pairs are independent of interactions used. 

The $12^+$ state in $^{52}$Fe (four-hole system) has been observed to be a spin trap~\cite{Ur98}, whose excitation energy is lower than that of $10^+_1$. This scheme can be reproduced nicely in large-scale $fp$ shell model calculations with realistic nucleon-nucleon interactions~\cite{Qi08}. In $0f_{7/2}$ shell, the energy of this unique state is given as,
\begin{equation}
E_{12}(^{52}{\rm Fe})=\frac{6}{13}\bar{V}_5+3\bar{V}_6+\frac{33}{13}\bar{V}_7,
\end{equation}
where $\bar{V}$ denotes hole-hole interaction matrix elements which can be extracted from the spectrum of $^{54}$Co~\cite{nudat}. We have a total number of three $10^+$ states (among which one with $T=1$). The energy of the first one is given as,
\begin{eqnarray}
\nonumber E_{10^+_1}(^{52}{\rm Fe})& =&0.310\bar{V}_3 + 1.429\bar{V}_4
\\  &&+ 0.497\bar{V}_5 + 1.571\bar{V}_6 + 2.193\bar{V}_7.
\end{eqnarray}
which is calculated to be lower than that of the $12^+$ state.
As a comparison, with two-body interactions extracted from the spectrum of $^{42}$Sc~\cite{Zam05,nudat}, the energy of the first $10^{+}$ state in $^{44}$Ti is calculated to be,
\begin{eqnarray}
\nonumber E_{10^+_1}(^{44}{\rm Ti})&=&0.245 V_3+1.473V_4\\
&&+0.570V_5+1.526V_6+2.186V_7.
\end{eqnarray}
The expansion coefficients of above two expressions similar to each other, which are obtained by diagonalizing the corresponding Hamiltonian matrices.

A similar spin trap, with $I^{\pi}=16^+$, has been predicted in the four-hole system of $^{96}$Cd below the doubly magic $^{100}$Sn~\cite{Ogawa}.
This prediction is supported by our $0g_{9/2}$-shell calculations with hole-hole interactions from Ref.~\cite{js} as well as large-space calculations in the $fpg$-shell. In $j=9/2$ shell, the energy of this state  is,
\begin{equation}
E_{16}(^{96}{\rm Cd})=\frac{8}{17}\bar{V}_7+3\bar{V}_8+\frac{43}{17}\bar{V}_9.
\end{equation}
We have a total number of three $14^{+}$ states. The energy of the first $14^{+}$ state is calculated to be,
\begin{eqnarray}
\nonumber E_{14^+_1}(^{96}{\rm Cd})&=&0.307\bar{V}_5+1.428\bar{V}_6\\
&&+0.493\bar{V}_7+1.572\bar{V}_8+2.200\bar{V}_9,
\end{eqnarray}
For the matrix elements we have $\bar{V}_5\approx \bar{V}_7$ and $\bar{V}_6\approx\bar{V}_8$~\cite{js,slg}. Thus the position of the $16^{+}$ state relative to the first $14^+$ state is sensitive to the strength of interaction matrix element $\bar{V}_9$ which is much more attractive than above mentioned elements. This observation has also been found in Ref.~\cite{Ogawa} with numerical calculations.

For $n=3$ and $|T_z|=1/2$ systems in a single-$j$ shell, similar to Eq.~(\ref{4pn}), the wave function of a state with total angular momentum $I$ can
be written as,
\begin{equation}
|\Psi_I\rangle=\sum_{J_{\alpha}} X_I(J_{\alpha}) | j^2(J_{\alpha})j;I \rangle,
\end{equation}
where $X_I(J_{\alpha})$ is the amplitude and $J_{\alpha}$ and $I$ are integers and half integers, respectively. In above equation, we have assumed that the system was decomposed into a nucleon pair and a odd nucleon. The Hamiltonian matrix can be constructed in a similar way to Eq.~(\ref{4h}).

In the system with  $n=3$, one has a total number of three pairs with different spins. For the number of pairs with a given $J$ one gets,
\begin{eqnarray}\label{num3}
\nonumber C^I_J&=&|X_I(J_{\alpha})|^2\delta_{J_{\alpha}J}\\
 &&+\sum_{J_{\alpha}J_{\alpha}'}X_I(J_{\alpha})N^I_J(J_{\alpha}J_{\alpha}')X_I(J_{\alpha}').
\end{eqnarray}
The symmetric matrix $N$ is given as,
\begin{eqnarray}
N^I_J(J_{\alpha}J_{\alpha}')=2\hat{J}_{\alpha}\hat{J}_{\alpha}'\hat{J}^2\left\{
\begin{array}{ccc}
 j & I & J \\
j & j & J_{\alpha} \\
\end{array}\right\}
\left\{\begin{array}{ccc}
j & I & J \\
j & j & J_{\alpha}' \\
\end{array}
\right\}.
\end{eqnarray}
As an illustration, in the lowe panel of Fig.~\ref{fig1} we gave calculations for coefficients $C^I_J$ along the yrast states of $^{43}$Sc and $^{43}$Ti. Calculations are done in the $0f_{7/2}$ shell with two-body interactions from Ref.~\cite{Zam05}.

For $I\neq j$, there is no $J=0$ pair. For $I=j$ one has $N^j_0(J_{\alpha}J_{\alpha}')=2J_{\alpha}J_{\alpha}'/\hat{j}^4$ and the expression for the number of $J=0$ pairs can be simplified as,
\begin{equation}
C^j_0=|X_j(0)|^2+\frac{2}{\hat{j}^4}\sum_{J_{\alpha}J_{\alpha}'}\hat{J}_{\alpha}\hat{J}_{\alpha}'X_j(J_{\alpha})X_j(J_{\alpha}').
\end{equation}
For the unique state with the maximum spin of $I=3j-1$, it can be easily seen from Eq.~(\ref{num3}) that we have $C^I_J=0$ for $J<2j-1$ and
\begin{equation}
C^{I=3j-1}_{J=2j-1}=C^{I=3j-1}_{J=2j}=\frac{3}{2}.
\end{equation}
The states with total spins $I=1/2$ and $I=3j-2$ are also unique. For the $I=3j-2$ state we have $C^I_J=0$ and $3/2$ for $J<2j-2$ and $J=2j-1$, respectively. For the numbers of $J=2j-2$ and $2j$ pairs, they are related to the $6j$ symbol as,
 \begin{equation}
C^{I=3j-2}_{J}=2(4j-1)(2J+1)\left|\left\{\begin{array}{ccc}
j &3j-2 & J \\
j & j & 2j-1 \\
\end{array}
\right\}\right|^2.
\end{equation}
As an example, the energy of the $I=17/2$ state in $j=7/2$ shell can be expanded as,
\begin{equation}
E_{17/2}(j=7/2)=\frac{19}{26}V_5+\frac{3}{2}V_6+\frac{10}{13}V_7.
\end{equation}
For the  $I=1/2$ state, the energy can be expressed as,
\begin{equation}
E_{1/2}(j)=\frac{3}{2}V_{j-1/2}+\frac{3}{2}V_{j+1/2}.
\end{equation}

The energy expressions for systems with three and four identical nucleons in a single-$j$ shell have been explored in textbooks~\cite{Talmi93} as well as in recent publications~\cite{Zam08,Esc06,Rowe01,Isa08,Zam07b,Talmi10}.
For four identical nucleons in a single-$j$ shell, the basis state can still be written as $|j^2(J_{\alpha})j^2(J_{\beta});I\rangle$ where it is assumed that the system was decomposed into two blocks. Since the Pauli principle is not acting, the overlap between such states is~\cite{Liotta81},
\begin{eqnarray}
\nonumber A^j_I(J_{\alpha}J_{\beta};J_{\alpha}'J_{\beta}')&=&\langle j^2(J_{\alpha})j^2(J_{\beta});I|j^2(J'_{\alpha})j^2(J'_{\beta});I\rangle\\
\nonumber &=&\delta_{J_{\alpha}J_{\alpha}'}\delta_{J_{\beta}J_{\beta}'}+(-1)^{I}\delta_{J_{\alpha}J_{\beta}'}\delta_{J_{\beta}J_{\alpha}'}\\
 && -4\hat{J}_{\alpha}\hat{J}_{\beta}\hat{J}_{\alpha}'\hat{J}_{\beta}'
\left\{
\begin{array}{ccc}
j&j&J_{\alpha}\\
j&j&J_{\beta}\\
J_{\alpha}'&J_{\beta}'&I
\end{array}
\right\},
\end{eqnarray}
where $J_{\alpha}$ and $J_{\beta}$ are even numbers.
For $I=0$ one obtains,
\begin{equation}
A^j_0(J_{\alpha}J_{\alpha};J_{\alpha}'J_{\alpha}')=2\delta_{J_{\alpha}J_{\alpha}'}+4\hat{J_{\alpha}}\hat{J_{\alpha}'}\left\{\begin{array}{ccc}
j&j&J_{\alpha}\\
j&j&J_{\alpha}'
\end{array}\right\},
\end{equation}
where $A^j_0(00;00)=(4j-2)/(2j+1)$.
In orbits with $j\leq7/2$, the dimension of spin zero states is $D^2_0(j)=1$. We found that the energy of this state can be written as,
\begin{equation}
E_0(j)=\sum_{J,{\rm even}}A^j_0(JJ;JJ)V_J.
\end{equation}

Similarly, the basis states of systems with three identical nucleon can be given as the coupling of a nucleon pair and a single nucleon as $| j^2(J_{\alpha})j;I \rangle$~\cite{Blom84}. These bases are not orthonormal to each other.
The overlap between different states is given as,
\begin{eqnarray}\label{3ov}
\nonumber B^j_I(J_{\alpha}J_{\alpha}')&=&\langle j^2(J_{\alpha})j;I|j^2(J'_{\alpha})j;I\rangle\\
 &=&\delta_{J_{\alpha}J_{\alpha}'}+2\hat{J}_{\alpha}\hat{J}_{\alpha}'
\left\{
\begin{array}{ccc}
j&j&J_{\alpha}'\\
j&I&J_{\alpha}\\
\end{array}
\right\}.
\end{eqnarray}
It is easily seen that for $I=j$ one has $B^j_j(J_{\alpha}J'_{\alpha})=1/2A_0(J_{\alpha}J_{\alpha}J'_{\alpha}J'_{\alpha})$.
Similar to systems with four identical nucleons, for $j\leq7/2$ we have,
\begin{equation}
E_j(j)=\sum_{J_{\alpha}}B^j_j(J_{\alpha}J_{\alpha})V_{J_{\alpha}}.
\end{equation}

The dimension of $J = j$
states in three-identical-nucleon system equals to the dimension of $J = 0$
states in four-identical-nucleon system, which is given as
$D=[(2j+3)/6]$. It equals to the rank of overlap matrices $A$ and $B$. The dimensions of matrices
$A$ and $B$ are the same, i.e., $n=(2j+1)/2$.

Seniority remains a good quantum number for systems with $j\leq7/2$~\cite{Talmi93}. For $9/2\leq j\leq 13/2$, there are two $I=j$ and $I=0$ states in three-identical-nucleon and four-identical-nucleon systems, respectively. In this case for interactions to conserve seniority, it has to satisfy one (or $[(2j-3)/6]$) linearly relationship~\cite{Talmi93,Rowe01,Isa08}.
In a  $n=3$ system, the conservation of seniority implies that $\langle\nu=1;j|\hat{V}|\nu=3;j\rangle$=0~\cite{Talmi93}.
Based on the overlap matrix, we found that the conservation condition can be written as,
\begin{equation}
\sum_{J>0}B^j_j(0J)\left(B^j_j(0J)-\frac{B^j_j(00)}{B^j_j(20)}B^j_j(2J)\right)V_J=0.
\end{equation}
For interaction $V_0$, the corresponding coefficient is zero. The same conservation condition can be got by employing the overlap matrix $A$ of the four-body system. This relation is sufficient to ensure that seniority is conserved for all states in shells with $9/2\leq j\leq 13/2$.
If seniority is conserved in the interaction, we found that the wave function of unique $\nu=1$, $I=j$ state for three-identical-nucleon system in any single-$j$ shell satisfies the relation,
\begin{equation}
\langle\nu=1;j|j^2(J_{\alpha})j;j\rangle=\frac{B^j_j(0J_{\alpha})}{\sqrt{B^j_j(00)}}.
\end{equation}
It implies that the wave function can be written as,
\begin{equation}
|\Psi_j(\nu=1)\rangle=\frac{1}{\sqrt{B^j_j(00)}}\sum_{J_{\alpha}}B^j_j(0J_{\alpha})|j^2(J_{\alpha})j;j\rangle.
\end{equation}
The wave function of the $\nu=0$, $I=0$ state in four-identical-nucleon system can be constructed in the same way.

In summary, in this work the number of nucleon pairs for three- and four-particle systems in a single-$j$ shell has been evaluated with the decomposition of the angular momentum in the framework of the multistep shell model. We have derived analytic expressions that are valid for all spins and isospins. Possible simplified relations are also explored with the symmetric properties of the angular momentum coupling coefficients.
With the expressions thus derived we explored the number of pairs in isomeric states as well as other unique (dimension one) states. The decomposition can also help us in understanding the conservation of seniority in single-$j$ shells.

Within the framework of the multistep shell model~\cite{Liotta81}, where the building blocks are different nucleon pairs, this analysis of pair numbers
can be extended to systems with many particles. These may help us to understand the property of $N\approx Z$ nuclei in single-$j$ shells as well as the structure of isomeric states
in many-particle nuclear systems, e.g., the spin-trap isomers in $^{51}$Fe and $^{95}$Pd. Further work in this direction is underway.

We thank R. J. Liotta (Stockholm) for stimulating discussions and careful reading of the manuscript. This work is motivated by a recent experimental proposal to study the isoscalar coupling scheme in $N=Z$ nuclei~\cite{Ced10}. The author is supported by a grant from the Swedish Research Council (VR).

\end{document}